\documentclass[aps,prb,onecolumn,citeautoscript,superscriptaddress,nofootinbib,eqsecnum,notitlepage,11pt]{revtex4-1}  
\pdfoutput=1

\usepackage{amsmath,amssymb,bm} 
\usepackage{graphicx}
\usepackage[tight]{subfigure}    
\usepackage{color} 
\usepackage[papersize={8.5in,11in}]{geometry}
\usepackage[colorlinks=true]{hyperref}  
\hypersetup{
    bookmarks=true,         % show bookmarks bar?
    unicode=false,          % non-Latin characters 
    pdftoolbar=true,        % show Acrobat
    pdfmenubar=true,        % show Acrobat 
    pdffitwindow=false,     % window fit to page when opened
    pdfstartview={FitH},    % fits the width of the page to the window
    pdftitle={My title},    % title
    pdfauthor={Author},     % author
    pdfsubject={Subject},   % subject of the document
    pdfcreator={Creator},   % creator of the document
    pdfproducer={Producer}, % producer of the document
    pdfkeywords={keyword1} {key2} {key3}, % list of keywords
    pdfnewwindow=true,      % links in new window
    colorlinks=true,       % false: boxed links; true: colored links
    linkcolor=magenta, %red,          % color of internal links (change box color with linkbordercolor)
    citecolor=blue,        % color of links to bibliography
    filecolor=magenta,      % color of file links
    urlcolor=blue           % color of external links
} 

\usepackage{amsfonts}
\usepackage{upgreek}
\usepackage{slashed}
\usepackage{latexsym}

\usepackage{epsfig}

\newcommand{\ing}{\includegraphics}
\newcommand{\bib}{\bibitem}
\newcommand{\ben}{\begin{equation}}
\newcommand{\een}{\end{equation}}
\newcommand{\bea}{\begin{eqnarray}}
\newcommand{\eea}{\end{eqnarray}}

\def\vx{{\vec{x}}}

\def\vphi{{\vec{\phi}}}
\def\vpsi{{\vec{\psi}}}

\newcommand{\nn}{\nonumber \\}

\renewcommand{\vec}[1]{{\mathbf{#1}}}
\newcommand{\ok}{\Omega_k}
\newcommand{\okd}{\dot{\Omega}_k}
\newcommand{\okdd}{\ddot{\Omega}_k}
\newcommand{\dt}{\Delta t}

\begin{document}

\title{Order parameter dynamics of the non-linear sigma model in the large $N$ limit}

\author{Sebastian Gemsheim}
\affiliation{Fakult\"at Physik, Technische Universit\"at Dresden, 01069 Dresden, Germany
}

\author{Ipsita Mandal}
\affiliation{Laboratory of Atomic And Solid State Physics, Cornell
University, Ithaca, NY 14853, USA}
\affiliation{Faculty of Science and Technology, University of Stavanger, 4036 Stavanger, Norway}

\author{Krishnendu Sengupta}
\affiliation{School of Physical Sciences, Indian Association for the
Cultivation of Science, Jadavpur, Kolkata-700032, India}

\author{Zhiqiang Wang}
\affiliation{Department of Physics and Astronomy, McMaster
University, Hamilton, Ontario, L8S 4M1, Canada}
\date{\today}

\begin{abstract}

We study non-equilibrium order parameter dynamics of the non-linear
sigma model in the large $N$ limit, using Keldysh formalism. We
provide a scheme for obtaining stable numerical solutions of the
Keldysh saddle point equations, and use them to study the order parameter
dynamics of the model either following a ramp, or in the presence of
a periodic drive. We find that the transient dynamics of the order
parameter in the presence of a periodic drive is controlled by the
drive frequency displaying the phenomenon of synchronization. We
also study the approach of the order parameter to its steady state
value following a ramp and find out the effective temperature of the
steady state. We chart out the steady state temperature of the
ordered phase as a function of ramp time and amplitude, and discuss
the relation of our results to experimentally realizable spin models.

\end{abstract}

\maketitle

\tableofcontents
%%%%%%%%%%%%%%%%%%%%%%%%%%%%%%%%%%%%%%
\section{Introduction}
\label{intro}

The study of quantum field theoretic systems driven out of
equilibrium has received a lot of interest, both in the context of
high energy physics and ultracold atom systems \cite{sdas1,sdas2,
sengupta1, polkovnikov1, rigol1}. There are several reasons for such
a surge of interest in this field. The first is the recent
experimental realizability of isolated quantum systems using
suitable combination of lasers and ultracold atoms. Such a setup may
emulate strongly correlated condensed matter models isolated from
its environment. The low energy properties of such correlated
systems are often described by quantum field theories. Furthermore,
the parameters of these ultracold atom systems are easily tunable;
therefore they serve as perfect test beds for studying quantum
dynamics of these models. The study of quantum field theories driven
out of equilibrium is therefore expected to provide relevant input
for understanding such experimental systems. The second motivation
is more theoretical in nature and involves developing an
understanding of non-equilibrium dynamics of quantum field theoretic
systems. An example of such endeavor involves the study of the rate
of excitation production in such systems due to the presence of the
drive. Such studies are particularly interesting near the critical
points of these models and in the presence of a linear or a periodic
drive. The former type of drives leads to the well-known
Kibble-Zurek (KZ) scaling \cite{kz1,anatoli1,ks1}, and the latter
leads to a realization of St{\"u}ckleberg interference phenomenon in
these systems \cite{stu1, ks2}. In addition, the study of such system
allows us to address the notion of universality in such
out-of-equilibrium systems, whose details may differ significantly
from their equilibrium counterpart \cite{rg1,ks3}.

Moreover, there are two other broad theoretical motivations for
studying such dynamics in field theoretic models. The first of them
involves understanding the transient dynamics of a driven field
theory. Such dynamics can in principle be complicated due to
the interplay of the drive frequency with several inherent frequency
scales of the model (arising out of its mass and interaction
parameters). The second involves the approach of such a driven system to
its steady state and eventual onset of thermalization. It is usually
expected that for a non-integrable model, such a steady state would
be thermal and shall thus be characterized by an effective
temperature \cite{rigol1}. These issues have been studied recently
in the context of the SYK model \cite{syk1}. However, the dependence
of this effective temperature on the drive amplitude and frequency
is a model dependent phenomenon, and has not been widely studied for
non-integrable quantum field theories beyond $d=1$ \cite{cardyref1}.

The study of out-of equilibrium dynamics in interacting field
theories, however, is technically challenging even at zero
temperature. The difficulty involved stems from the fact that
the properties of the driven system may depend, in principle, on all its
states. This situation is to be contrasted with the study of equilibrium
field theories, where one requires the knowledge of the ground state
of the theory, which can be computed using several known perturbative
techniques. For this reason, most of the studies on dynamics of such
field theories have concentrated on free field theories
which are integrable. Such studies are interesting in their own
right and lead to a wealth of information about dynamical aspects of
several systems \cite{ft1}. However, several properties of such
integrable field theories are fundamentally different from their
non-integrable counterparts. One key example involves the property
of steady states that the driven system reaches at sufficiently long
time; for integrable theories, such states are not necessarily
thermal and are described by a generalized Gibbs ensemble (GGE). In
contrast, the steady state of driven non-integrable models are
usually thermal in accordance with the eigenstate thermalization
hypothesis (ETH) \cite{ethlit}. However a detailed study of such
steady state behavior for $d>1$ quantum field theoretical systems
turns out to be difficult.

In this work, we carry out such a study for quantum rotors described
by the non-linear sigma model in the large $N$ limit \cite{subir1}. This
model serves as an effective description for several spin systems
\cite{sref1}. Moreover, it is conjectured to be dual to higher spin
gauge theories in $AdS_4$ \cite{hspinref1}. It is well known that such
higher spin gauge theories are often intractable; thus we expect the
study of non-equilibrium dynamics of the more tractable large $N$
vector model to provide useful information about several dynamic
properties of its dual counterpart. The dynamics of the paramagnetic
phase of this model has been studied in Ref.\ \onlinecite{sdas3}. Here, we
concentrate on the ordered phase of the model, and study the behavior
of its order parameter, either following a ramp or in the presence of
a periodic drive. Such a drive or ramp is implemented by making the coupling
parameter of the model time-dependent. In this work, we always
restrict ourselves to the case where we are within the ordered phase
at all times, and are sufficiently away from the critical point.

The main results of our work are as follows. First, we set the
Keldysh saddle point equations for the driven model and provide a
prescription for obtaining stable numerical solutions of these
equations. We find that our numerical method leads to stable
convergent solutions for the order parameter dynamics as long as the
drive or ramp amplitude is sufficiently small. Second, using this method, we
study the long time steady state behavior of the order parameter
following a ramp. We study the approach of the order parameter to
its steady state value and compute the effective temperature $ {\mathcal{T}}_{\mathrm{eff}} $ of the
steady state. We chart out this effective temperature as a function
of the ramp time and amplitude. Third, we study the transient
dynamics of the order parameter in the presence of a periodic drive.
We find that the transient dynamics is controlled by the drive
frequency and the order parameter oscillation displays
synchronization. We explain the reason for such synchronization
using the large $N$ Keldysh saddle point equation of motion.
Finally, we discuss our main results and point out their
experimental implications.

The plan of the rest of the paper is as follows. In Sec.\
\ref{form}, we study the dynamics of the large $N$ non-linear sigma
model using Keldysh formalism and chart out our numerical method.
This is followed by Sec.\ \ref{res} where we present our main
results for periodic drive and quench dynamics. Finally, we discuss
our main results and conclude in Sec.\ \ref{diss}.

\section{General formalism}
\label{form}

In this section, we first set up the Keldysh formalism following the
treatment of the paramagnetic phase of the model in Ref.\ \onlinecite{sdas3}, and
obtain the saddle point equations of the order parameter in Sec.\
\ref{ferro}. This is followed by a prescription for an efficient
numerical solution of these saddle point equations in Sec.\
\ref{numsol}.

%%%%%%%%%%%%%%%%%%%%%%%%%%%%%%%%%%%%%%%%%%%%%%%%%%%%
\subsection{Quench dynamics in the ferromagnetic phase}
\label{ferro}

The action of the large $N$ non-linear sigma model in equilibrium is
given by \cite{subir1}
\begin{eqnarray}
S[\phi^{\ast}, \phi] = \int d^dx\, dt  \left[ \frac{N}{2 \,g(t)}
(\partial_\mu \vphi (\vx,t))\cdot(\partial^\mu \vphi (\vx,t)) +
\lambda (\vx,t) \left( \vphi \cdot \vphi - 1 \right ) \right], \label{ac1}
\end{eqnarray}
where $\vphi (\vx,t)$ is an $N$ dimensional vector with real
components. In the large $N$ limit, $N \rightarrow \infty$ while
$g(t)$ remains finite ($O(1)$). The field $\lambda(\vx,t)$ is a
Lagrange multiplier which imposes the constraint $\vphi (\vx,t)
\cdot \vphi (\vx,t) =1$. The saddle point equation of the model
implements this constraint on the average, and leads to a solution
with uniform $\lambda$. It is well known that the critical coupling
at equilibrium is given by:
\begin{eqnarray}
\frac{1} {g_c({\mathcal{T}}) } &=& \int \frac{d^dk} {(2 \pi)^d}  \frac{ \coth
\left ( \beta \, k \right )}   {2 \, k}\,,  \label{eqgc}
\end{eqnarray}
where $k= |\vec k|$, $\mathcal{T}$ is the temperature, $\beta=1/{\mathcal{T}}$, and we have set the velocity $c=1$.

In this work, we are going to concentrate on the magnetically
ordered phase which occurs at $g<g_c$.  In this phase, we write
$\phi= (\rho, \Pi_1,\Pi_2, \cdots , \Pi_{N-1} )$ to allow for a
finite expectation value $\rho$ of one of the components of the
vector field $\phi$: $\langle \phi_1 \rangle=\rho$. In equilibrium,
the magnetization $\rho$ is a constant; however, a time-dependent
$g(t)$ is expected to lead to a time-dependence of $\rho$. Also, we note
that we shall discuss global protocols in this work, which allow
$\rho$ to be independent of space. The action in the ordered phase in
the presence of the drive is then given by
\ben S = \int d^dx\, dt
\left[ \frac{N}{2\, g(t)} \Big \lbrace (\partial_\mu \Pi
(\vx,t))\cdot(\partial^\mu \Pi (\vx,t)) + (\partial_t \rho (t))^2
\Big \rbrace + \lambda (\vx,t) \left ( \Pi \cdot \Pi + \rho^2 - 1
\right ) \right].
 \label{f-1} 
\een where $\Pi (\vx,t) =(\Pi_1,\Pi_2, \cdots , \Pi_{N-1} ) $ is an $(N-1)$-dimensional vector with real components. In the large $N$ limit, $N \rightarrow \infty$ with
$g(t)$ remaining $O(1)$. Redefining the fields as 
\ben \Pi \rightarrow \vpsi = \sqrt{\frac{N}{g(t)}} \,
\Pi\,,
\een
 the action becomes 
 \ben S[\psi^{\ast},\psi] = \int dt {\mathcal
L}, \quad {\mathcal L}= \frac{1}{2} (\partial \vpsi)^2 + \frac{N}{2 g(t)}\dot \rho^2
-\frac{1}{2} \Sigma (\vx,t) \left( \vpsi^2  + (\rho^2 -1) \frac{N
}{g(t)} \right) + \frac{ N \alpha(t)}{g(t)}  \,, 
\label{f-2} 
\een
where 
\ben \alpha (t) = \frac{1}{4} \left[ \frac{3}{2}\left(
\frac{\dot{g}}{g}
  \right)^2 - \left( \frac{\ddot{g}}{g} \right) \right],
~~~~~~~~ -\frac{1}{2} \Sigma(\vx,t) = \frac{g(t)}{N}\lambda(\vx,t) +
\alpha (t) \,. 
\label{f-3} 
\een 
The last term in Eq.~(\ref{f-2}) is
field independent, and can be therefore ignored. We note that we
have ignored all total derivative terms in writing the expression
for ${\mathcal L}$.

Following Ref.\ \onlinecite{sdas3}, we express the partition function
using the Schwinger-Keldysh path integral technique. This involves
defining the fields $\psi_{+(-)}$ on the forward and backward
contours. The partition function can be then written as
\begin{align}
{\mathcal Z} = \int  {\mathcal D} \vpsi_{\pm} {\mathcal D}
\Sigma_{\pm}~e^{i \left[S(\vpsi_+,\Sigma_+) -
S(\vpsi_-,\Sigma_-)\right] } ,
 \label{keldysh}
 \end{align}
 where $S[\psi_{\pm}, \Sigma_{\pm}] \equiv
 S_{\pm}$ are given by Eq.\ (\ref{f-3}), and are defined on the
 forward and backward Keldysh contours. Next, integrating out the fields
$\vpsi_\pm$ leads to the effective action for $\Sigma_\pm$ and
$\rho_{\pm}$ as
\begin{align}
S_{\mathrm{eff}} &= \frac{(N-1)}{2}{\rm Tr}~\log (D^{-1}) + N  \int
d^dx \,dt \, \frac{ \dot \rho_+^2 -  \dot \rho_-^2 } {2 g(t)}
%\nonumber\\&&
- N\int d^2x \,dt \, \frac{   \left ( 1-\rho^2_+(t)
\right )\, \Sigma_+ - \left ( 1-\rho^2_- (t) \right )\,\Sigma_- }
{2g(t)}  \,,
\end{align}
where $D$ is the propagator matrix whose inverse is \ben D^{-1}
= \left(
\begin {array}{cc}
\partial^2-\Sigma_+&0\\
\noalign{\medskip} 0& -\partial^2 + \Sigma_-
\end {array}
\right) \,.
\een
%%%%%%%%%%%%%%%%%%%%%
In the large $N$ limit, saddle point equations therefore take the
form:
\begin{eqnarray}
&& \frac{1- \rho^2_+ (t) }{g(t)} =- {\rm Tr}~D_{++} \,
,~~~~~~~\frac{1 - \rho^2_- (t) }  {g(t)} = {\rm
  Tr}~(D_{--}) \,, \nn
%%%%%%%%%%%%%%%%%%%%%%%%%%%%%
&& \rho_+ \, \Sigma_+ = \ddot \rho_+ \,, \quad \rho_- \, \Sigma_- =
\ddot \rho_- \,, \label{1..6}
\end{eqnarray}
where $D^{-1}_{ ++} =
\partial^2-\Sigma_+ $ and $D^{-1}_{ -- } =- \partial^2 +
\Sigma_-$.
%%%%%%%%%%%%%%%%%%%
At the saddle point, we should have $ \rho_{+}=\rho_- = \rho$ and $
\Sigma_{+}= \Sigma_- = \Sigma$, similar to the structure in the
paramagnetic phase. This means we need to solve the two coupled
equations:
\begin{eqnarray}
&& 1- \rho^2 (t) = g(t) \, {\rm Tr}~\mathcal D\, , \quad
%%%%%%%%%%%%%%%%%%%%%%%%%%%%%
\rho(t) \, \Sigma(t) = \ddot \rho(t)  \,, \nn && \mathcal D =
-D_{++} = D_{ -- }  \,. \label{coupled-eqn}
\end{eqnarray}

Following Ref.\ \onlinecite{sdas3}, the first of the two coupled equations can
be written as:
\begin{eqnarray}
\rho^2(t) + g(t) \int \frac{d^d k}{(2\pi)^d} \frac{1}{2 \,
\Omega_k(t)} \coth\left (\frac{\beta  \,  k}{2} \right ) &=& 1\,,
\label{1-13}
\end{eqnarray}
where $\Omega_k(t)$ satisfies the equation:
\ben
\frac{1}{2}\frac{\ddot {\Omega}_k}{\Omega_k} - \frac{3}{4} \left(
\frac{\dot{\Omega}_k}{\Omega_k} \right)^2 + \Omega_k^2 = k^2 +
\Sigma(t)\,. 
\label{1-12} \een
%%%%%%%%%%%%%
In equilibrium, we have $\dot \rho =0$. Therefore, the saddle has
$\Sigma_+ = \Sigma_- \equiv  0 $ for, say, when $g=$constant in the
ordered phase with non-zero $\rho$, and the other equation to solve
is: \ben
  \rho^2 +  g  \,\int d^dx ~ \langle \vx,t | D | \vx,t \rangle_\beta  = 1 \,.
\label{1-7} \een However, here we seek a solution such that $g(t) $,
and hence $\Omega_k(t)$, are slowly-varying functions of $t$. Let us
assume an expansion
\begin{eqnarray}
\rho(t) &=& \sum \limits_{n=0}^{\infty} \epsilon^n
\,\rho^{(n)}(t)\,,\quad \Omega_k(t) = \sum \limits_{n=0}^{\infty}
\epsilon^n \, \Omega_k^{(n)}(t) \,.
\end{eqnarray}
%%%%%%%%%%
The initial conditions are: $ \Sigma(0)=0$ so that $\rho(0 )= \sqrt{ 1-
g(0)/g_c({\mathcal{T}},t=0)}$, and $ \dot \rho (0) = 0$. We note that a
time-dependent $g$ generates a non-zero $\Sigma(t)$, which is not
necessarily small. Nevertheless, for protocols with small $\omega$,
$\dot \Sigma$ is expected to be small [${\rm O}(\epsilon)$]. This
implies that $\dot \Omega_k$ and $\ddot \Omega_k $ are ${\rm O}
(\epsilon)$ and ${\rm O}(\epsilon^2)$, respectively. It is only under this
condition that one can consider the derivative terms in $
\Omega_k(t) $ as higher order. Performing an expansion and
collecting all terms with the same order in $\epsilon$, we find, at
zeroth order,
\begin{eqnarray}
\Omega_k^{(0)} (t) &=&  \sqrt{ k^2 + \Sigma(t)}, \, \,
\dot{\Omega}_k^{(0)} = \frac{\dot \Sigma} {2 \sqrt{ k^2 + \Sigma }},
\,\,  \ddot{\Omega}_k^{(0)}  = \frac{\ddot \Sigma}  {2 \sqrt{ k^2 +
\Sigma }} -  \frac{\dot \Sigma^2}  {4 \,( k^2 + \Sigma )^{3/2}}.
\label{perteq1}
\end{eqnarray}
Then using these in Eq. \ \ref{1-12} at next order, we get:
\begin{equation}
\frac{1}{\Omega_k}  = \frac{1}{\sqrt{k^2 +\Sigma^2 }} \Big[ 1+
\frac{\ddot \Sigma}{ 8 \,(k^2 +\Sigma)^2} -\frac{5 \, \dot\Sigma^2}
{32  \,(k^2 +\Sigma)^3 } \Big]\,. \label{zeroth0}
\end{equation}
Thus one can develop a systematic perturbative expansion of the
saddle point equations for small $\omega$. We shall, however, be
interested in behavior of the system beyond this regime. Thus in
Sec.\ \ref{numsol} we develop a prescription for exact numerical
solution of the saddle point equations.

%%%%%%%%%%%%%%%%%%%%%%%%%%%%%%%%%%%%%%%%%%%%%%%%%%%%%%%%%%%%%%%%%
\subsection{Numerical solution of the saddle point equations}
\label{numsol}

In this section, we provide a numerically efficient prescription to
solve Eqs.\ (\ref{1-13}) and (\ref{1-12}). In what follows, we shall
focus on $d=3$, for which the ordered phase exists at finite
temperatures. We first consider the initial condition: $\dot{\rho} =0,
\Sigma =0 $ at $t=0$. From the discussion regarding the zeroth order
solution (before Eq.~(\ref{zeroth0})), it becomes clear that one has
$\Omega_k (0)= k$ and $\dot{\Omega}_k (0) = \frac{\dot{\Sigma}(0) }
{2 \,k}$. At this point, we note two essential points. First, for
numerical solution of these equations, it is useful to have an
initial condition for $\dot{\Sigma}(0)$ which we shall take to be a
small initial value. The qualitative features of our numerical results would not depend on
the precise value of $\dot \Sigma(0)$.
Second, since we are in $d=3$, the
coupling has the dimension of mass. 
For numerical calculations, we rescale all
length and time scales by $\sqrt{g_c(0)}$, where $g_c(0)$ is the zero temperature critical coupling value at equilibrium
and hereafter we will set $g_c(0)=1$.

%%%%%%%%%%%%%%%%%%%%%%%%%
We start with a zero initial temperature where the system is in its
ground state. One can then write a set of three self-consistent
equations that we need to solve. These are given by
\begin{eqnarray}
    &&\rho^2(t) + g(t) \int \frac{d^3 k}{(2\pi)^3} \frac{1}{2\,\ok(t)} = 1   \,, \nn
    &&\frac{1}{2} \frac{\okdd}{\ok} - \frac{3}{4} \left( \frac{\okd}{\ok} \right)^2 + \ok^2
    = k^2 + \frac{\ddot{\rho}(t)}{\rho(t)}  \,,  \nn
    && g(t) = g_0 + \left (g_1-g_0 \right ) \zeta(t)\,,
\label{eqorigin}
\end{eqnarray}
where $g_0 = g(t=0)$, and $\zeta(t)$ is a function of time that satisfies $\zeta(0)=0$. The form of the second equation suggests
the introduction of new variables:
\begin{eqnarray}
    \rho(t) \equiv \rho(0) \, e^{D(t)} \,,  \quad
    \Omega_k(t) \equiv k \,e^{B_k(t)}\,,
\end{eqnarray}
with initial conditions $D(0)=B_k(0)=0$. Taking derivatives, we get:
\begin{eqnarray}
    \ddot{\rho}(t) &=& \rho(t)  \left[ \ddot{D}(t) + \dot{D}^2(t) \right]  ,
    \quad \dot{\Omega}_k(t) = \Omega_k(t) \dot{B}_k(t),   \quad
    \ddot{\Omega}_k(t) = \Omega_k(t) \left[ \ddot{B}_k(t) + \dot{B}_k^2(t)   \right] .
\label{eqnew}
\end{eqnarray}
Using these, the second equation of Eq.~(\ref{eqorigin}) transforms
into (omitting time arguments for simplicity)
\begin{equation}
    \ddot{B}_k - \frac{1}{2} \dot{B}_k^2  + 2k^2 \left( e^{2B_k} - 1 \right) = 2\left[ \ddot{D} + \dot{D}^2 \right] \,.
\end{equation}
We transform the above second order ordinary differential equation
(ODE) to first order ODEs, by introduction of a new set of variables,
namely,
\begin{eqnarray}
    \chi_k &=&  \dot{B}_k \,,\quad \nu = \dot{D}   \,.
\end{eqnarray}
The new equation takes the form:
\begin{equation}
    \dot{\chi}_k - \frac{\chi^2_k}{2}  + 2\,k^2 \left( e^{2\, B_k} - 1 \right) = 2\left[ \dot{\nu} + \nu^2 \right].
\end{equation}

Using a simple central difference scheme with discretized time and step
size $\dt$:
\begin{equation}
    f'(n) = \frac{f(n+1) - f(n-1)}{2\dt}  + \mathcal{O}\left( \dt^2 \right ),
\end{equation}
the ODEs read (the time arguments are shifted for convenience):
\begin{eqnarray}
    B_k(n+1) - B_k(n-1) &=& 2\,\dt \, \chi_k(n) \,\nn
    D(n+1) - D(n-1) &=& 2\dt\, \nu(n)\,,\label{diffeq1} \\
      \chi_k(n) - \chi_k(n-2) - \dt\left [\chi_k^2(n-1) + 4 k^2 \left( 1 - e^{2\,B_k\,(n-1)} \right)  \right ] &=& 2\,\nu(n) - 2\,\nu(n-2) + 4\,\dt\,\nu^2\,(n-1)\,.
    \nonumber
\end{eqnarray}
In order to fulfill the constraint equation, we seek an equation of
the form $B_k(n, D(n))$. Inserting the second expressions for
$\nu(n)$ into the last equation of Eq.\ (\ref{diffeq1}), we get:
\begin{eqnarray}
    \chi_k(n) &=&  \chi_k(n-2) + \dt\, \chi_k^2(n-1) + 4\,\dt\,k^2 \left( 1 - e^{2B_k(n-1)}  \right) \nonumber\\
    && + 2\,\frac{D(n+1) - D(n-1)}{2\dt} - 2 \, \nu \,(n-2) + 4\,\dt\,\nu^2\, (n-1)\,.\label{chieqn}
\end{eqnarray}
Using Eqs.\ (\ref{diffeq1}) and (\ref{chieqn}), for $B_k(n)$, we obtain:
\begin{eqnarray}
    B_k(n) &=& \left[ B_k(n-2) + 2\, D(n) - 2\,D(n-2) \right] + 2\,\dt
    \left[  \chi_k\, (n-3)  - 2\,\nu\, (n-3)   \right] \nn
    && \quad + 2\, \dt^2   \left[ \chi_k^2\,(n-2) +  4\,k^2 \left( 1 - e^{2B_k(n-2)}  \right)  + 4\,\nu^2\,(n-2)
    \right]. \label{bneq}
\end{eqnarray}
Note that the term $\propto \dt^2k^2$ in Eq.\ (\ref{bneq}) is small
for sufficiently small $\dt$ and for the k modes
which participate in the dynamics till the time we carry out the numerics.

The equation for the constraint can be written in terms of $\rho(n)$
as
\begin{eqnarray}
\rho^4(n)- \rho^2(n)+ \tilde{G} &=& 0\,,  
\quad \tilde G =
\frac{g(n)}{(2\,\pi)^3} \int_0^{\Lambda} d^3{k}\, \frac{\rho^2(0)}{2\,k}\, e^{2 D(n) - B_k(n,
D(n))} \,,
\label{coneq1}
\end{eqnarray}
where $\Lambda=\sqrt{8\pi^2}$ is the ultraviolet cutoff.
The last equation can be solved symbolically in terms of $\tilde G$.
It yields:
\begin{equation}
    \rho^2(n) = \frac{1}{2} \left[ 1 \pm \sqrt{1 - 4\,\tilde{G}} \right]. \label{coneq2}
\end{equation}
Hence, the integration for $\tilde{G}$ has to be performed only once per
time step. The sign of the solution is chosen such that
$\Delta\rho(n) \equiv | \rho(n)-\rho(n-1) |$ is minimized. The
condition $4\,\tilde{G}\leq1$ must be fulfilled at all times. This
restricts the validity of this approach to quenches or drives which
keep the system well within the ordered phase.

The initial conditions
$\left( \ddot{\rho}(0)=\dot{\rho}(0)=\ddot{\Omega}_k(0)=\dot{\Omega}_k(0)=0  \right ) $
are chosen to be:
\begin{eqnarray}
    \rho_0 &=& \sqrt{ 1 - \frac{g(0)}{g_c} } \,,\quad
    \chi_k(0) = \chi_k(1) = 0 \,, \quad
    \nu(0) = \nu(1) = 0 \,,\nn
    B_k(2) &=& B_k(1) = B_k(0) = 0 \,,\quad
    D(2) = D(1) = D(0) = 0      \,.
\end{eqnarray}
The full problem has been reduced to a simple numerical integration
and iteration per time step. This procedure can be computed very
fast and at a low computational cost. In the next section, we shall
use this computational procedure to study the dynamics of the order
parameter in the presence of a periodic drive or following a sudden
quench of $g$. We shall restrict our study here within the ordered
phase for which $g(t)/g_c \le 1/2$ at all times during the
evolution; we have checked numerically that this is enough to ensure
the stability of the above-mentioned procedure.

\section{Numerical Results}
\label{res}

In this section, we chart out our numerical results, which involve
two separate classes of studies. The first involves steady state of
the driven system, while the second pertains to short-time transient
dynamics.
%%%%%%%%%%%%%%%%%
\begin{figure}
\centering {\ing[width=0.45\linewidth]{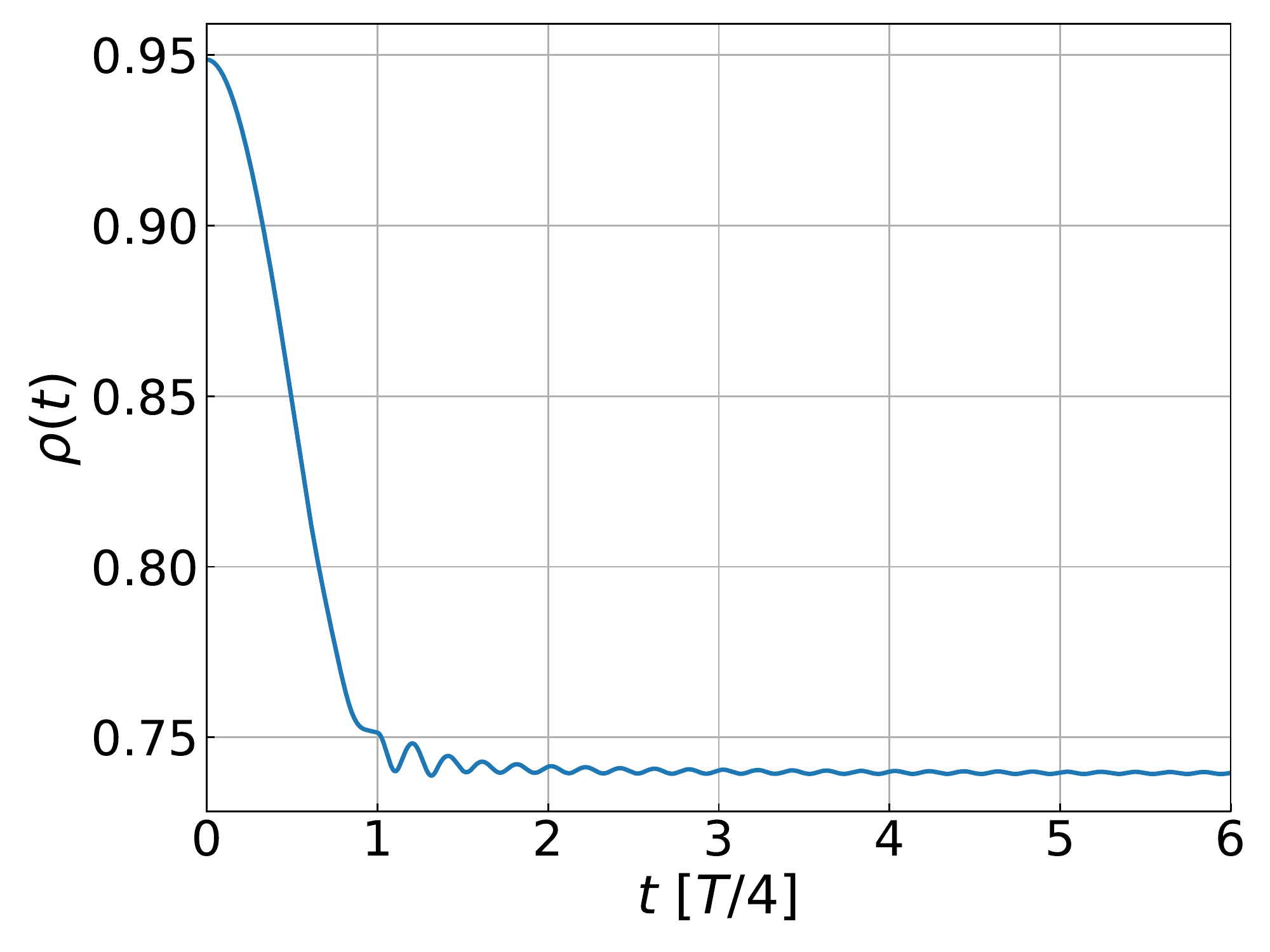}} 
\caption{Plot of the
order parameter $\rho(t)$ as a function of $t/T$, showing approach of
$\rho$ to its steady state value for $T=4$, $g_i/g_c=0.1$ and $g_f/g_c= 0.45$.
See text for details.} 
\label{fig1a}
\end{figure}
%%%%%%%%%%%%%%%%%%%%%%%%%%%%%%%

For addressing the steady state of this model, we use the following
protocol. We drive the system using a drive protocol with time
period $T$:
\begin{eqnarray}
g(t)= g_i +(g_f-g_i) \, \sin^2 \left(2 \, \pi\, t/T \right )  
\label{gprotocol}
\end{eqnarray}
for $t \leq T/4 \,,$ and then let the system evolve with a time independent
Hamiltonian $H[g_f]$. This amounts to Hamiltonian evolution of the
system following a ramp with a characteristic time $T$. One expects
the system to reach a steady state in the course of such an evolution, and
the goal of our study is to understand the behavior of the
magnetization in this steady state. Here we choose the initial
($g_i$) and the final ($g_f$) values of the coupling such that $g(t)
\le g_c/2$ for all times; this ensures numerical stability as
discussed in Sec.\ \ref{numsol}. We track the evolution of the order
parameter (magnetization) $\rho(t)$ during its subsequent evolution
following the ramp. The behavior of $\rho(t)$, for $T=4$ and
$g_i(g_f)=0.1 (0.45) g_c \, ,$ is shown in Fig. \ref{fig1a}. We find that
$\rho(t)$ shows a fast decay (within $t \le T/4$), and then displays
small oscillations around a steady state value. The amplitude of
these oscillations decay with $t$, and for $t\ge 10$, the
magnetization reaches its steady state value. Thus we find a
relatively fast onset of steady state behavior for order parameter
dynamics of this model in its ordered phase. This behavior is to be
contrasted with that in the paramagnetic phase, where such fast onset
was not observed \cite{sdas3}.

\begin{figure}
\centering {\ing[width=0.5\linewidth]{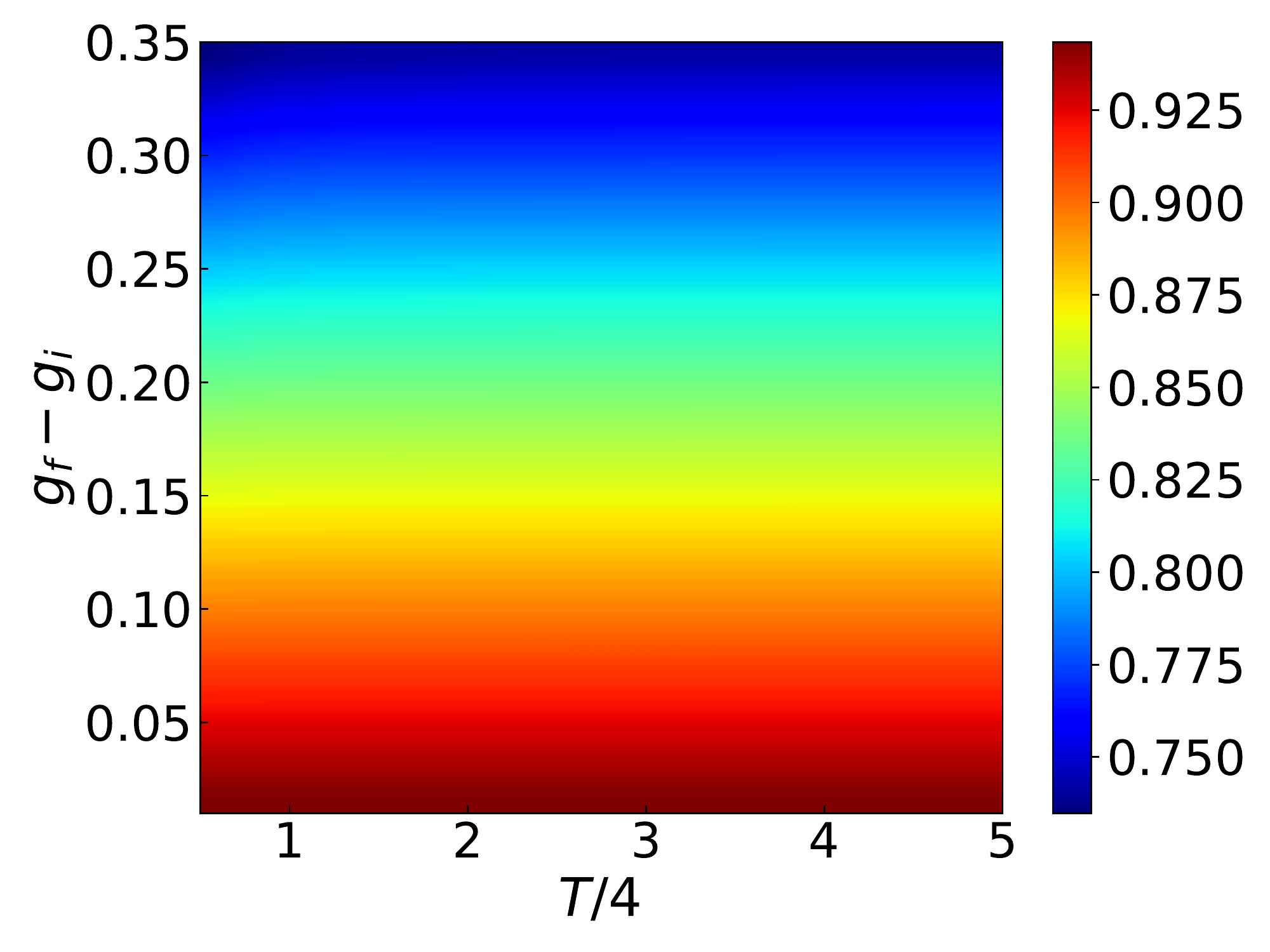}} \caption{Plot of the
steady state value of $\rho$ as a function of the drive amplitude
$(g_f-g_i)/g_c$ and ramp time $T$, for $g_i/g_c=0.1$. The plot shows
$\rho$ to be a monotonically decreasing function of the drive
amplitude for any $T$ in the ordered phase. See text for details.}
\label{fig2a}
\end{figure}

\begin{figure}
\centering {\ing[width=0.45\linewidth]{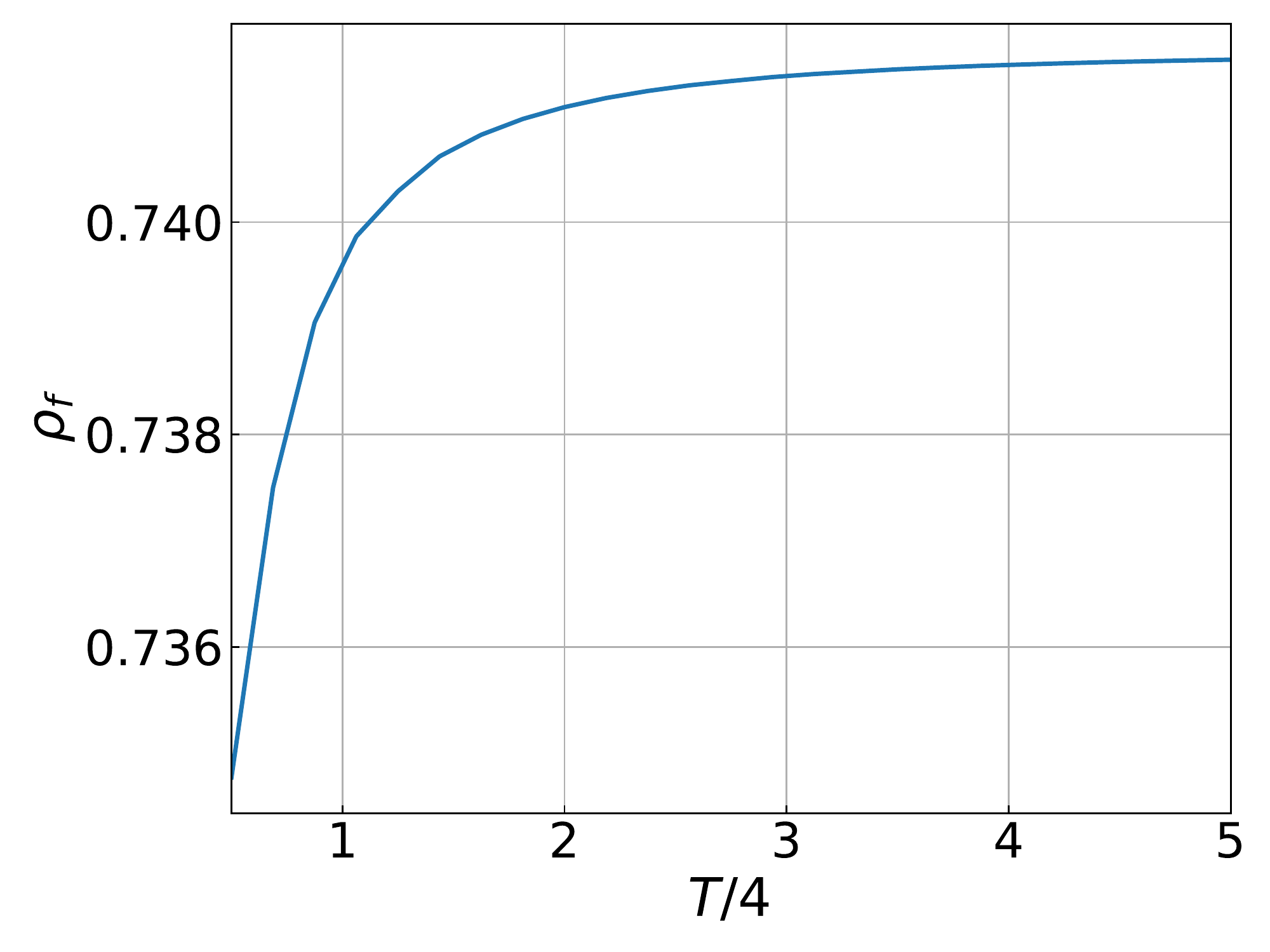}} \caption{ Plot of
the steady state value of $\rho$ as a function of the ramp time $T$,
for $g_f/g_c= 0.45$ and $g_i/g_c=0.1$ (fixed). The plot shows that
$\rho$ plateaus at large $T$. See text for details.}
\label{fig3a}
\end{figure}

In Fig.\ \ref{fig2a}, we plot the steady state order parameter value
as a function of the ramp amplitude $ \left( g_f-g_i \right ) $ and the time period
$T$, for $g_i/g_c=0.1$. We find that the steady state order parameter
value is a monotonically decreasing function of the ramp amplitude.
This behavior may be understood by noting that for any fixed $T$, a
larger ramp amplitude amounts to pumping more energy in the system.
Consequently, the steady state, which is a thermal state for
non-integrable models, exhibits higher effective temperature for
larger ramp amplitude. Both a larger $g_f$ and higher effective temperature lead to a lower steady state value of
$\rho$. In contrast, we find that for a fixed $g_f-g_i$, the steady
state value of $\rho$ is almost a constant within the range of $T$
studied. This feature is specifically pointed out in Fig.\
\ref{fig3a}, where we find that the steady state value of $\rho$
plateaus to a constant value with increasing $T$. This is a
consequence of the gapped nature of the system, leading to almost no
energy absorption at low drive frequencies (large $T$).

\begin{figure}
\centering {\ing[width=0.45\linewidth]{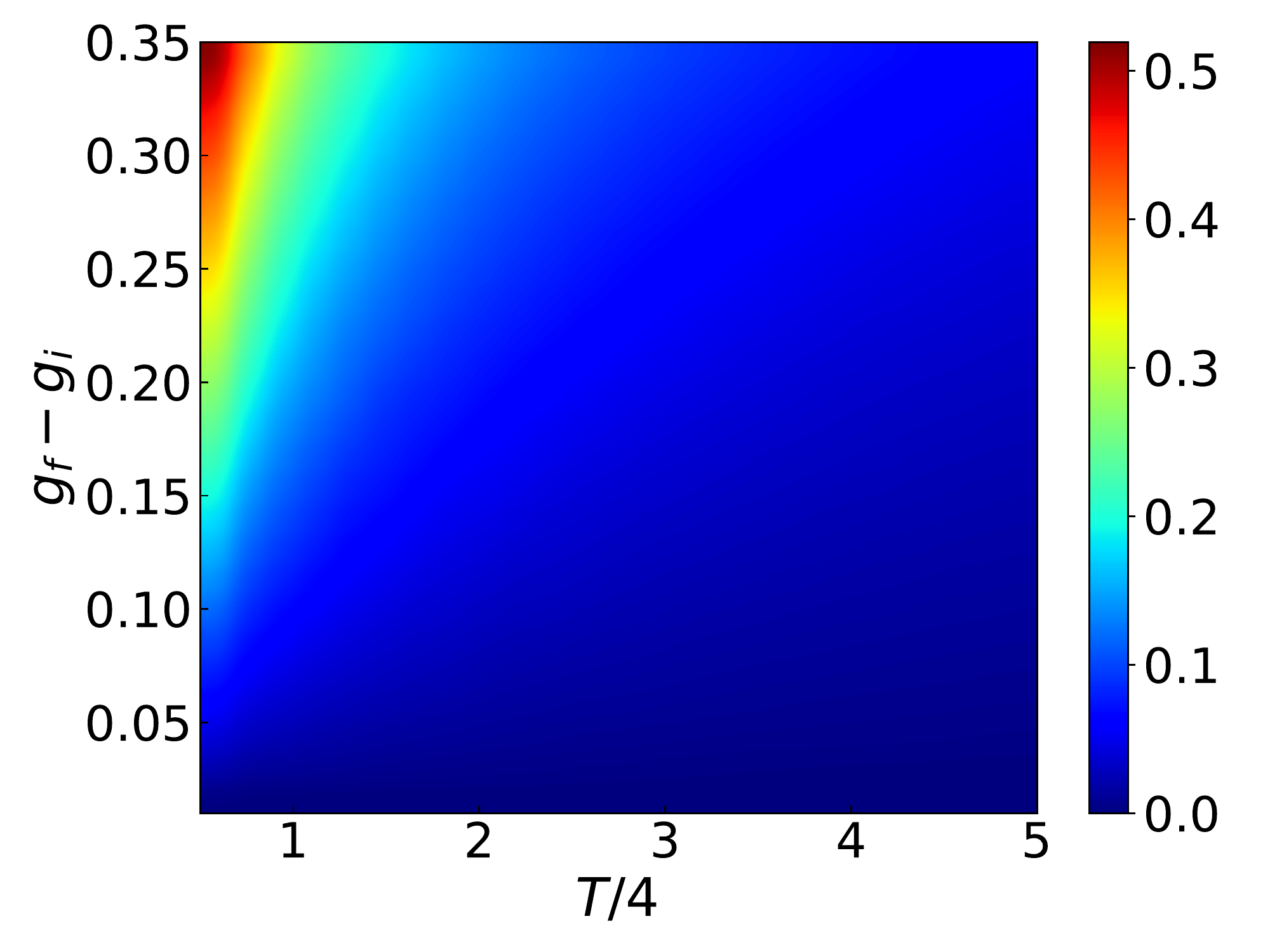}} \caption{ Plot of the
effective temperature of the steady state as a function of $T$ and
$(g_f-g_i)/g_c$, for a fixed $g_i/g_c=0.1$. The plot shows that the
effective temperature increases with decreasing $T$ and increasing
drive amplitude. See text for details.}
\label{fig4a}
\end{figure}

Next, in Fig.\ \ref{fig4a}, we plot the effective temperature $ {\mathcal{T}}_{\mathrm{eff}}$ of the
steady state as a function of the ramp amplitude and $T$. $ {\mathcal{T}}_{\mathrm{eff}}$ can be calculated from
$\rho_f^2=1-g_f/g_c( {\mathcal{T}}_{\mathrm{eff}})\,,$ with $\rho_f$ being the steady state value of $\rho$, and
$g_c( {\mathcal{T}}_{\mathrm{eff}})$ is obtained from Eq.\ (\ref{eqgc}). 
We find
that ${\mathcal{T}}_{\mathrm{eff}}$ increases with
both decreasing $T$ and large ramp amplitude. This behavior is
expected since for gapped closed systems, efficient energy
absorption can not occur at $T \ge [\Delta_0(g_i)]^{-1}$, where
$\Delta_0$ denotes the zero temperature equilibrium energy gap for
$g=g_i$.

\begin{figure}
\centering {\ing[width=0.45\linewidth]{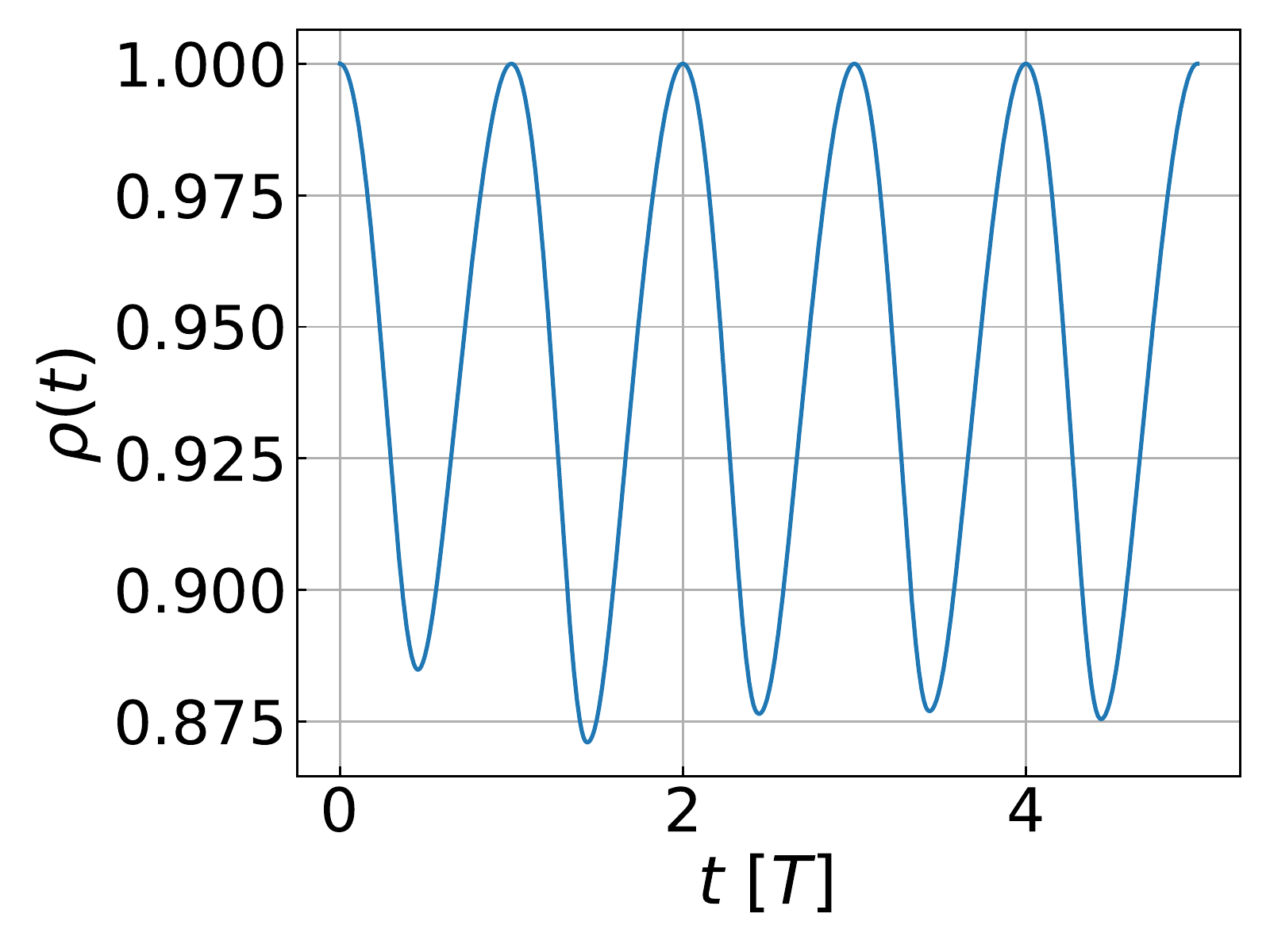}} \hspace{1.0 cm}
\centering
{\ing[width=0.48\linewidth]{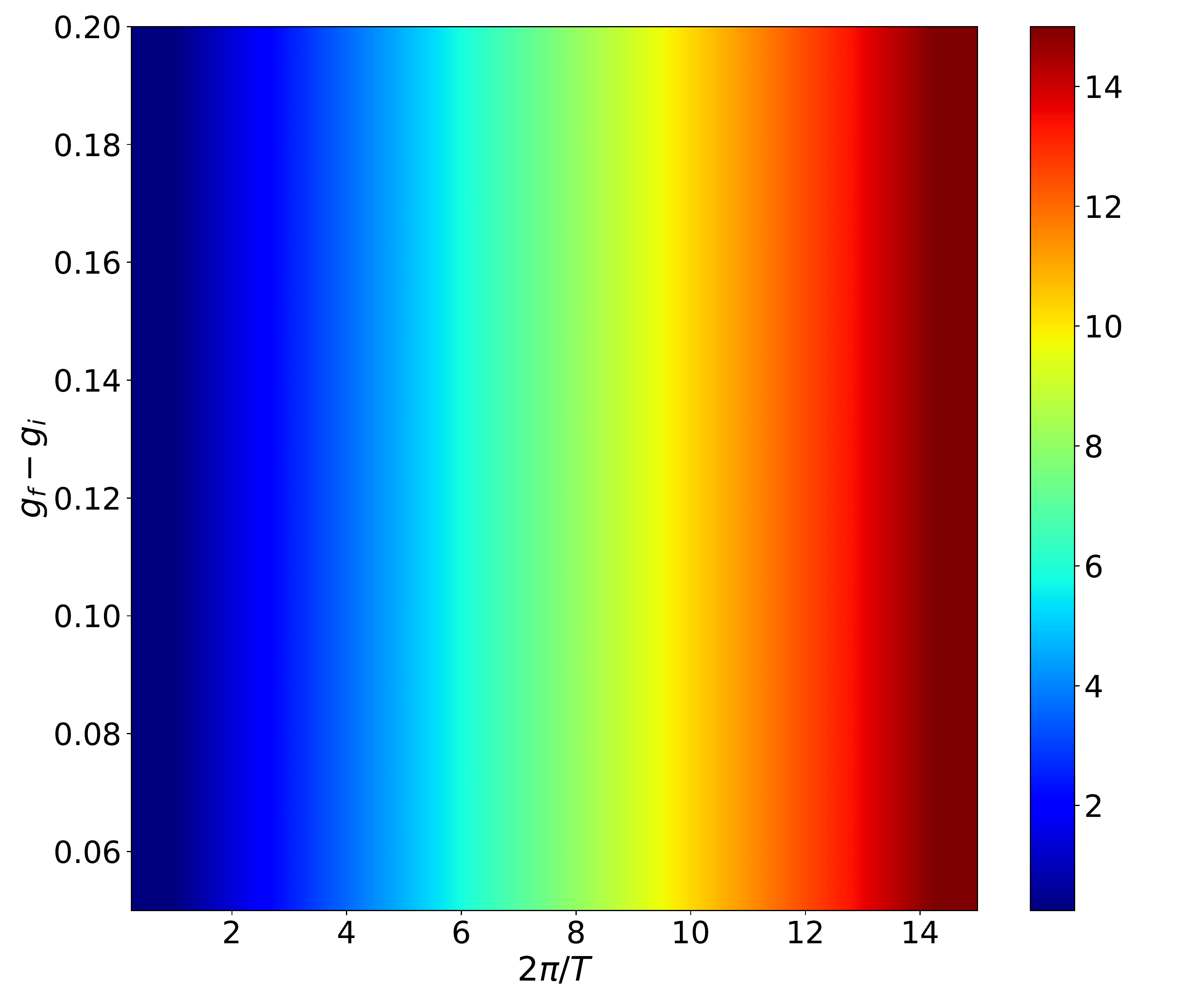}} \caption{Left: Plot of the
$\rho(t)$ as a function of $t/T$, for $ (g_f-g_i)  /g_c =0.2$,
$g_i/g_c=0 $ and $T \in [2\,\pi/ 15, 4 \cdot 2\,\pi]$. The plot shows transient oscillations
of $\rho$. Right: Plot of the oscillation frequency of $\rho$ as a
function of drive amplitude and $2\, \pi/T$, indicating almost
perfect synchronization. See text for details.} \label{fig5a}
\end{figure}

Finally, we discuss the transient dynamics of the model. For this, we
drive the system using the protocol given by Eq.\ (\ref{gprotocol}),
and track the dynamics of the order parameter in the presence of the
drive. The oscillation of the magnetization $\rho(t)$ is plotted in
the left panel of Fig.\ \ref{fig5a} as a function of time. The right
panel of Fig.\ \ref{fig5a} indicates the plot of the oscillation
frequency of $\rho$ as a function of both drive amplitude and
frequency. From these plots, we find that the oscillations have the
same time period as the drive for a wide range of drive amplitudes
and frequencies. This synchronization can be understood from our
saddle point solution in the following manner. First, we note from
Eq.\ (\ref{coneq2}), the dynamics of $\rho(t)$ is controlled by
$\tilde G$. Next, from Eq.\ (\ref{coneq1}), we note that $\tilde G \sim
g(t)$ at any instant. Moreover, from Eq.\ (\ref{bneq}), it is easy to
see that the exponent $\left[ 2\,D(n) - B(n,D(n)) \right ] $, which appears in $\tilde
G$ within the integral is always small. This follows directly from
Eq.\ (\ref{bneq}), the boundary condition $B(0)=0$, and the fact that
$\Delta t$ is always a small quantity. Consequently, the time
dependence of $\tilde G$ is always controlled by $g$. Moreover, since
we scale all quantities by the cutoff $\Lambda$, it is easy to see that
$\tilde G \ll 1$ for any $g(t)$. As a result, one can expand
$\sqrt{1-4 \,\tilde G} \simeq 1-2\, \tilde G + O(\tilde G^2)$. Thus,
from Eq.\ (\ref{coneq2}), one finds that the time dependence of $\rho$
is essentially the same as that of $\tilde G$. Thus, synchronization
occurs in this model as a structure of its saddle point equations,
which governs the dynamics.

\section{Discussion}
\label{diss}

In this work, we have studied the magnetization dynamics in the
ordered phase of the non-linear sigma model within a large $N$
approximation, in the presence of either a ramp or a periodic drive. Our analysis uses Keldysh path integral techniques, and obtains the saddle point
equations describing the magnetization dynamics. A key result of
this work is to provide a scheme to obtain stable numerical solution
of these saddle point equations. The method we chart out is
effective in the ordered phase of the model as long as $g(t)/g_c <
1/2$. We find that any further proximity to the critical points
leads to numerical instability. Thus the dynamics of magnetization
in such systems near the critical point still remains an open
question which we leave for future study.

Using this method, we obtain several results regarding both
transient dynamics and long-time steady states of the order
parameter. For the former, we find that the oscillation of the
magnetization synchronizes with the time period $T$  of the drive
for all ranges of drive amplitudes. We tie this behavior to the
structure of our saddle point equation, and show that this phenomenon
is expected to be robust in the ordered phase. Such a robustness is
confirmed in our numerical simulation for a wide range of $T$.

We also study the behavior of the steady states that the system
reaches via evolution, subsequent to a ramp with characteristic time
$T$. We find that the steady state is always thermal, and that the
onset of thermalization occurs within a small value of of $t/T$. We
compute the temperature of the steady state as a function of both the ramp time and the amplitude. We find that the steady state
temperature increases with increasing ramp amplitude; in contrast,
within the range of ramp times we study, it displays a relatively weak
dependence on the ramp time $T$. The latter behavior can be
attributed to the presence of a gapped spectrum in the ordered phase
of the system.

There are several spin models which can be described under various
approximations by such non-linear sigma model \cite{sref1, subir1},
within the large $N$ approximation. Such systems often have a large
effective $S$ due to spin-orbit coupling, which makes them ideal
candidates for large $N$ analysis. Our analysis indicates that the
magnetization dynamics of such systems in the ordered phase should
show fast approach to a thermal steady state for evolution, following
a ramp. Moreover, the transient order parameter dynamics in the
presence of a periodic drive would display synchronization.

To conclude, we have used the Keldysh technique to study non-equilibrium
magnetization dynamics in the ordered phase of the non-linear sigma
model within the large $N$ approximation. We have studied both transient
dynamics of the order parameter in the presence of periodic drive,
and its approach to a steady state following a ramp. Our results
indicate that the transient order parameter oscillation synchronizes
with the drive frequency. Moreover, the system evolves to a thermal
steady state following a ramp, whose temperature is charted out as a
function of the ramp time and the amplitude.

%%%%%%%%%%%%%%%%%%%%%%
\section{Acknowledgments}

We thank Subhodip Saha for participating in the initial stages of the project.

\section{Author contribution statement}
The author names in this paper are alphabetically ordered. All authors have equal contributions.

%%%%%%%%%%%%%%%%%%%%%%%%%%%%%%%%%%%%%%%

\end{document}